# A Deep-learning Real-time Bias Correction Method for Significant Wave Height Forecasts in the Western North Pacific

**Wei Zhang**
Ocean University of China
Qingdao, 266100, China
weizhang@ouc.edu.cn

**Yu Sun**
Ocean University of China
Qingdao, 266100, China

**Yapeng Wu**
Ocean University of China
Qingdao, 266100, China

**Junyu Dong**
Ocean University of China
Qingdao, 266100, China
dongjunyu@ouc.edu.cn

**Xiaojiang Song**
National Marine Environment
Forecasting Center
Beijing, 100082, China
xjsong@nmefc.cn

**Zhiyi Gao**
National Marine Environment
Forecasting Center
Beijing, 100082, China

**Renbo Pang**
National Marine Environment
Forecasting Center
Beijing, 100082, China

**Boyu Guoan**
National Marine Environment
Forecasting Center
Beijing, 100082, China

## ABSTRACT

Significant wave height (SWH) is one of the most important parameters characterizing ocean waves, and accurate numerical ocean wave forecasting is crucial for coastal protection and shipping. However, due to the randomness and nonlinearity of the wind fields that generate ocean waves and the complex interaction between wave and wind fields, current forecasts of numerical ocean waves have biases. In this study, a spatiotemporal deep-learning method was employed to correct gridded SWH forecasts from the European Centre for Medium-Range Weather Forecasting System Integrated Forecast System Global Model (ECMWF-IFS). This method was built on the trajectory gated recurrent unit deep neural network, and it conducts real-time rolling correction for the 0–240-h SWH forecasts from ECMWF-IFS. The correction model is co-driven by wave and wind fields, providing better results than those based on wave fields alone. A novel pixel-switch loss function was developed. The pixel-switch loss function can dynamically fine-tune the pre-trained correction model, focusing on pixels with large biases in SWH forecasts. According to the seasonal characteristics of SWH, four correction models were constructed separately, for spring, summer, autumn, and winter. The experimental results show that, compared with the original ECMWF SWH predictions, the correction was most effective in spring, when the mean absolute error decreased by 12.972–46.237%. Although winter had the worst performance, the mean absolute error decreased by 13.794–38.953%. The corrected results improved the original ECMWF SWH forecasts under both normal and extreme weather conditions, indicating that our SWH correction model is robust and generalizable.

## 1 Introduction

Surface winds and waves control energy exchange between the atmosphere and the oceans [1, 2]. Therefore, accurate forecasting of ocean waves is key to assessing climate change and exploiting wave energy [3, 4, 5].The Intergovernmental Panel on Climate Change report indicated that wave height, an important parameter characterizing ocean waves, plays a crucial role in estimating the impacts of climate change, such as coastal erosion, storm generation and development paths, and increased storm frequency and intensity [6]. In addition, long-term forecasting of ocean wave heights is of great significance due to its practical applications related to shipping, military activities, and coastal interests. The Western North Pacific (WNP) region is affected by frequent human activities and frequent complex atmospheric processes such as monsoons, subtropical highs, and tropical cyclones [7, 8]. Complex atmospheric conditions lead to large biases in ocean wave height forecasts. Therefore, real-time and rapid correction of short-term and long-term gridded forecast data from the European Centre for Medium-Range Weather Forecasting (ECMWF) for significant wave height (SWH) in this region is necessary.

As a big-data method, machine learning has been widely used in recent years to correct biases in numerical forecast model results. In 2021, automatic bias correction of the SWAN wave model was implemented using a Bayesian algorithm [9]. In the same year, researchers proposed that machine-learning models could be used to improve the accuracy of classical numerical models' wave estimation. Multi-layer perception, gradient-boosted decision trees (GBDTs), and an ensemble method were used to improve the model. This method allowed prediction of four ocean parameters: SWH, average wave period, peak period, and average onset direction [10]. Traditional machine-learning methods are often used for small study areas, such as lakes and rivers. Gan et al. [11] used Light Gradient Boosting Machine to predict water levels in the lower Columbia River. Notably, the GBDT algorithm has been improved and extended while remaining widely used; in 2016, the eXtreme Gradient Boosting (XGBoost) algorithm was proposed, which adds a regularization component to the loss function of GBDT that considers the complexity of the model, thereby simplifying the model and preventing overfitting [12]. Experiments show that XGBoost has marked advantages over other traditional machine-learning algorithms such as random forest and GBDT [13]. Therefore, XGBoost is also widely used for correction and forecasting of ocean parameters such as sea-surface wind fields and wave heights. This study used XGBoost as a baseline to compare with the correction performance of the proposed deep-learning methods.

The traditional machine-learning forecast correction described above generally focuses on small-scale lakes and rivers. However, with a large research area such as the WNP and large-scale complex grid data, integrated learning methods such as XGBoost have lower efficiency and accuracy than those of deep-learning models. Compared with deep neural networks, ensemble learning is more suitable for small datasets and may yield poor results for complex data such as images or sequences [14]. Deep learning is suitable for processing large-scale data and extracting features and information, and its effectiveness in improving predictions of ocean features has been demonstrated. Specifically, a sequence-to-sequence network that was used to reconstruct missing ocean wave features using a Bayesian parameter optimization strategy achieved good results [14]. Later, the sequence-to-sequence network proposed in that study was applied to renewable-energy prediction and again achieved better results than traditional methods [15].

Work on marine parameter prediction or correction is described as follows. Recently, a deep-learning method was proposed for predicting sea surface height anomalies (SSHAs), which uses long short-term memory (LSTM) to build a time-series learning network to solve the long-term dependence problem in the prediction task [16]. Wang et al.[17] used a fully connected neural network to correct the effective wave height and wind-speed variables of the HY2B altimeter, which effectively reduced the error between the two estimates, and identified the best input parameter combination for the effective wave height correction model. This paper employs the popular deep-learning framework trajectory gated recurrent unit (TrajGRU) [18] for correction of SWH grid forecast data. To reduce the number of correction models and increase the number of training samples, we use a rolling correction strategy (see Section 3.1.2 for details). To achieve real-time correction, the only network input is forecast data from ECMWF, excluding delayed data types such as reanalysis data. Overall, the smart-grid correction model developed based on TrajGRU with a traditional loss function can achieve rapid correction at different time points. This function effectively reduces the systematic error of the model, with the mean absolute error (MAE) of the near-term forecast (0–24 h) reduced by 30.02–33.407% and the MAE of the forward forecast (24–240 h) reduced by 12.286–28.105%.

However, these results indicate room for improvement. Deep-learning architectures typically employ root mean square error (RMSE) and MAE as loss functions for regression tasks. The inherent averaging effect of such traditional loss functions can lead to large errors in some pixels in the system [19]. To reduce the number of such pixels, we designed a pixel-switch loss function and used it for pre-training fine-tuning. The pixel-switch loss function fine-tunes the correction model for pixels with large biases during pre-training. This pixel-level optimization process maximizes the accuracy of SWH correction (see Section 3.3 for details). The results show that the overall error of the output from the model corrected using the pixel-switch loss function was reduced by 2–8% compared with the model trained using the traditional loss function.

As the ECMWF provides forecast data for ocean parameters such as sea-surface wind fields, we propose two schemes for SWH correction. The first method is driven only by wave fields, while the second adds wind fields as a driver to the wave-field-driven method [20, 21]. The experimental results obtained from the two methods were compared and analyzed to identify the best correction model. The rest of this article is organized as follows. The data used in this study are described in Section 2, the methods are presented in Section 3, an analysis of the experimental results is included in Section 4, and the conclusions are presented in Section 5.

## 2 Study Area, Data and Proposed Model

### 2.1 Study area

The study area was the WNP region, with a spatial range of 0–45 °N and 100 °E–175 °E, as shown in Fig. 2. Notably, the corrected results for SWH may be mixed with land areas, and information contained in the land areas may negatively impact the model training process. In this study, we used static masks to avoid impacts on land regions (for more details, see Section 3.3).

### 2.2 Data

We used the fifth-generation ECMWF Reanalysis of Global Climate (ERA5)[1] data [22], which is an updated version of the European Medium-Range Weather Forecast Reanalysis Data Interim Version (ERA-Interim) data [23], as ground-truth data. ERA-Interim and ERA5 are global atmospheric and oceanic reanalysis data provided by the ECMWF, and ERA5 reanalysis data offer advantages over ERA-Interim data, especially in terms of spatial and temporal resolution [24]. Experimental results have shown that ERA5 SWH is more accurate than ERA-Interim SWH relative to independent buoy data [22]. The ERA5 reanalysis data are collected using a continuous data-assimilation scheme with a period of 1 h forward, assimilating a large number of observational data sources such as buoys and satellite altimeters in combination with improved humidity analysis and satellite data error correction technologies. ERA5 provides more accurate initial conditions for re-forecasting, leading to better forecast results [25]. Current ECMWF forecasts have low accuracy for certain weather phenomena, and ERA5 can provide guidance for determining the accuracy of the ECMWF forecasting products [26]. Therefore, we selected ERA5 data as the ground-truth data for this study. This dataset is updated daily with a delay of approximately 5 days; its spatial resolution is 0.5 ° and its temporal resolution 1 h.

All forecasts in the ECMWF model[2,3], are obtained using numerical models. The atmospheric model and data assimilation system developed by ECMWF are designated the integrated forecasting system (IFS) [27]. ECMWF-IFS updates its forecast daily at 00:00 UTC and 12:00 UTC, and updates the forecast data for the next 10 days each time. The temporal resolution of the wind-speed forecast data is 3 h for the first 6 days and 6 h for the last 4 days, and its spatial resolution is 0.1 ° × 0.1 °. The temporal resolution of the SWH forecast data is 6 h, and the spatial resolution is 0.25 ° × 0.25 ° covering the world's oceans.

To unify the spatial and temporal resolutions of the three data sources, we spatially interpolated the wind-speed forecast data and the ERA5 data onto a Cartesian grid with a resolution of 0.25 ° using a bilinear interpolation method. Temporally, the ECMWF wind-speed forecast data and ERA5 reanalysis data with a 6-h resolution were matched with the ECMWF SWH forecast data at the corresponding time.

These data were then divided into training, validation, and test sets, which are described in Section 3.1.2. Any two of the three sets were disjointed. In this study, the training and validation dataset sizes varied between 278,800 and 414,100 depending on the season and were sufficient for training reliable deep-learning models [28].

### 2.3 Proposed model

In this study, we propose a deep-learning bias correction method to correct SWH in the WNP region. The method consists of four parts: feature engineering, pre-training with the traditional loss function, fine-tuning with the pixel-switch loss function, and correction and model inference. Fig. 1 shows the overall architecture of the bias correction method developed in this study and the function of each component.

(A) Feature engineering. After obtaining SWH, wind-speed data, and ERA5 data from ECMWF, the spatiotemporal resolution of the original samples was unified, and the test sets were constructed according to the seasonal characteristics of SWH. As ECMWF releases forecasts for 10 days at a time, and our model is statically corrected for one time point, to reduce the number of models, a rolling correction method was employed to correct the forecast data globally for the 10-day period released by ECMWF at time $t$. Details of the feature engineering are described in Section 3.1. The rolling correction method is illustrated in Fig. 4.

---

[1]https://cds.climate.copernicus.eu/cdsapp#!/dataset/reanalysis-era5-single-levels?tab=overview

[2]https://www.ecmwf.int/en/forecasts/datasets/set-ii

[3]https://apps.ecmwf.int/codes/grib/param-db/?id=140229

(B) Pre-training using the traditional loss function. The spatiotemporal sequence model TrajGRU is designed to train using a training set and validate against a validation set, thereby obtaining the pre-trained optimal model. Detailed information about TrajGRU networks is provided in Section 3.2. The structure of TrajGRU is illustrated in Fig. 5 and the encoder–decoder in Fig. 6.

(C) Fine-tuning using the pixel-switch loss function. The pixel-switch loss function (see Section 3.3 for details) is used to fine-tune the optimal model identified in step (B) to obtain the final corrected product. Details of pre-training and fine-tuning are provided in Section 3.4.

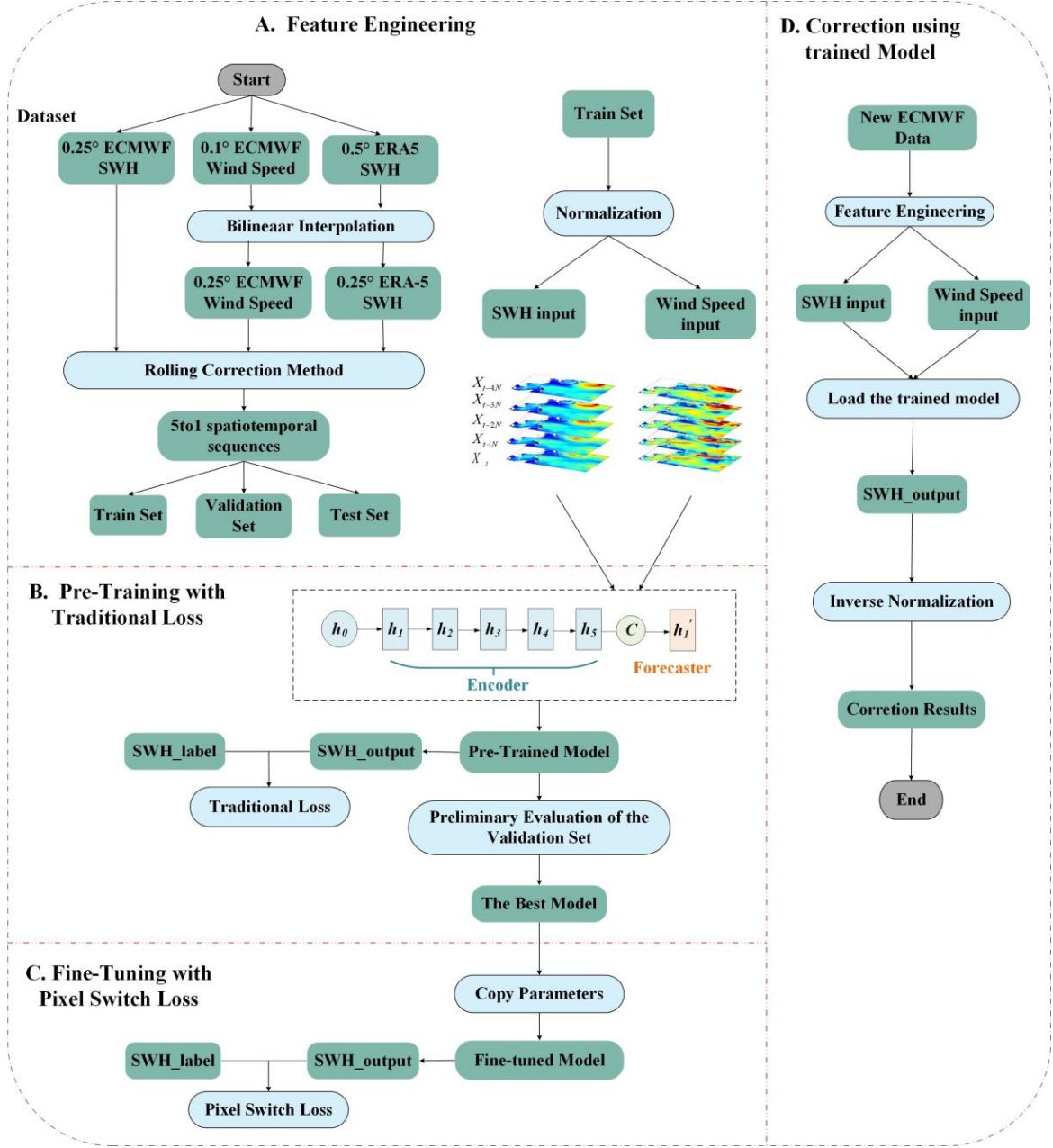

Figure 1: Framework of the proposed correction model: (A) Feature engineering. (B) Pre-training of the model using the traditional loss function and input of the 240-h ECMWF forecast into TrajGRU. (C) Fine-tuning of the model using the same input as the pre-training process and replacement of the loss function with the pixel-switch loss function proposed in this paper. (D) Correction and model inference.

(D) Correction using the trained model (called model inference in the context of machine learning). When the new ECMWF forecast data are released, sample features are constructed according to step (A) and input into the fine-tuned model in Step (C), and the corrected products are obtained directly. Notably, because the feature space obtained through sample construction does not contain ERA5 data, its availability is not delayed, and thus the model can be applied to real-time correction of SWH.

## 3 Related Methodology

### 3.1 Feature engineering

This section details the feature engineering for step (A) in Fig. 1. The reason for seasonal correction is explained in Section 3.1.1, and details about the rolling correction method are provided in Section 3.1.2.

#### 3.1.1 Construction of the training set and test set according to seasonal characteristics

Statistical analysis of Topex/Poseidon altimeter data over 75 consecutive months from October 1992 to December 1998 showed that the distribution of SWH in the Pacific Ocean exhibited marked seasonal changes and corresponded strongly to the distribution of wind speeds over the Pacific Ocean [29]. Therefore, we statistically analyzed the spatial and temporal distributions of SWH data from ERA5 in 2021 to construct the seasonal correction model. The statistical results are shown in Fig. 2 and Fig. 3.

As shown in Fig. 2, the difference in the distribution of SWH mean values among seasons indicates that SWH exhibits seasonal variation across the WNP region. Across nearly the entire WNP region, SWH was significantly higher in winter than in summer, with the largest differences of 3 $m$ or more appearing at middle to high latitudes of 30 °N–45 °N, 140 °E–175 °E. Seasonal differences in SWH are smaller near the equator than in the WNP. As shown in Fig. 3, the median and interquartile ranges of the four seasons significantly differ, such that median and maximum values are significantly higher in winter than in other seasons.

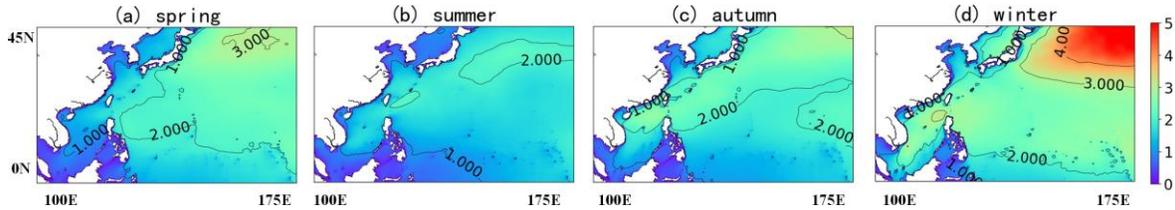

Figure 2: Average significant wave height in spring (a), summer (b), autumn (c), and winter (d).

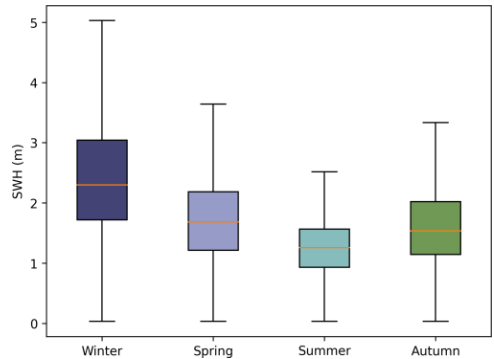

Figure 3: Differences in wave conditions for four seasons.

Due to the marked seasonal differences in SWH in the WNP, we conducted correction experiments for wave height among seasons. All data were divided into training, validation, and test sets. The training and validation sets for the spring corrected model were from March 2021, and the test set was from April 2021; the training and validation sets for the summer corrected model were from July 2021, and the test set was from August 2021; the training and validation sets for the autumn corrected model were from September 2021 and October 2021, and the test set was from November 2021; the training and validation sets for the winter corrected model were from December 2020 and January 2021, and the test set was from February 2021. After construction of the training and test sets, data in the training set were used for modeling, while data in the test set were used for evaluation and analysis of the experiments described in Section 4. ECMWF forecast data for the following 10 days were divided into two periods: days 1–4 and days 6–10. In this manner, only two models per season required training to correct the forecast data for the wind fields over the next 10 days.

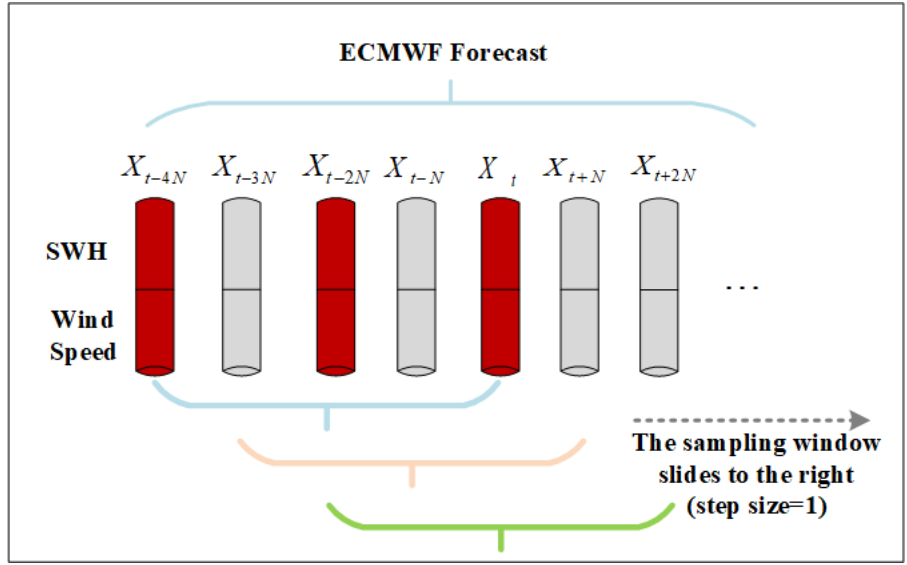

Figure 4: Schematic diagram of the rolling correction scheme. Each cylinder represents two 180 ×300 grids, with the red cylinder representing the issue time and the gray cylinder representing the lead time. For the input sequence $X$ in each sliding window, $N$ = 6 h.

### 3.1.2 Rolling Correction Method

In this section, we elaborate on the transformation of the problem of SWH bias correction into a spatiotemporal learning problem [30, 31, 32], which allowed us to predict $K$-frame images based on historical $S$-frame images. First, the data used in this study were all grid products. For example, the WNP SWH forecast data at time $t$ can be regarded as an $H \times W$ grid $X_t \in R^{H \times W}$. Each grid point can be regarded as a pixel in each frame of the SWH historical images, and the grid data for $S$ neighboring forecast moments can be regarded as an $S$-frame image sequence expressed as a three-dimensional tensor, namely $X = \{X_1, X_2, \cdots, X_s\}, X \in R^{S \times H \times W}$. We constructed a 5-to-1 model for spatiotemporal samples based on the SWH variable, using the time series length $S$ = 5 as input and $K$ = 1 as output.

The ECMWF provides SWH forecast data for the following 10 days. At each data issue time, the dataset covers a total of 41 forecast times over the following 10 days at 6-h intervals. To reduce the number of correction models and increase the training sample size, we propose a rolling correction scheme for ECMWF forecast data. As shown in Fig. 4, the size of the sampling window was set to 5, and it was rolled to the right with a step size of one forecast moment. For correction of the current time $t$, the ECMWF forecast data for five adjacent times were used as the input sequence $X = \{X_{t-4}, X_{t-3}, X_{t-2}, X_{t-1}, X_t\}, X \in R^{5 \times H \times W}$, the corrected sequence output by the model was $\hat{X} = \{\hat{X}_t\}, \hat{X} \in R^{1 \times H \times W}$, and the corresponding true value sequence of ERA5 was $Y = \{Y_t\}, Y \in R^{1 \times H \times W}$, where $H$ and $W$ are the height and width of the input image, respectively ($H$ = 180, $W$ = 300 in this study). The rolling correction method can expand the sample size to over 270,000, which is sufficient for training a reliable deep-learning model [28].

As noted in Section 3.1.1, the SWH distribution in the WNP has a strong correspondence with the wind-speed distribution. Therefore, to obtain SWH prediction data of the highest accuracy, we adopted the 10-$m$ $UV$ component released by ECMWF as an auxiliary feature for which we conducted feature engineering together with SWH variables after synthesizing wind speeds to ensure that the deep-learning model could obtain sufficient information. Four modeling schemes, as listed in Table 1, were compared in this study.

The XGBoost-S-TLoss scheme employs the XGBoost model and uses only the correction method of modeling SWH variables. TrajGRU-S-TLoss refers to the correction method that employs the spatiotemporal sequence model TrajGRU and uses only the SWH variable for modeling; its correction results were compared with the those of the XGBoost traditional machine-learning model. The input field of TrajGRU-SW-Tloss is driven by the SWH and wind-speed variables. The modified results of TrajGRU-S-TLoss were used to explore the best input combination for the SWH correction model. TrajGRU-SW-PLoss employed a correction method based on the pre-training and fine-tuning pixel-switch loss function proposed in this study (described in detail in Section 3.3). Based on the optimal model obtained at the pre-training stage, the pixel-switch loss function was used for fine-tuning, and the correction results of TrajGRU-SW-PLoss were compared with those of TrajGRU-SW-TLoss to validate the TrajGRU-SW-PLoss correction method.

In the following sections, we describe the construction of the corrected model and define the pixel-switch loss function in detail. Furthermore, XGBoost, a traditional machine-learning method that has been extensively used for prediction-correction tasks, is briefly introduced and compared with TrajGRU as a baseline.

Table 1: Configurations of the four models.

| Model | Method | Input of Model | Loss Function |
|---|---|---|---|
| XGBoost-S-TLoss | XGBoost | SWH | Traditional loss function |
| TrajGRU-S-TLoss | TrajGRU | SWH | Traditional loss function |
| TrajGRU-SW-TLoss | TrajGRU | SWH, Wind Speed | Traditional loss function |
| TrajGRU-SW-PLoss | TrajGRU | SWH, Wind Speed | Pixel Swich loss function |

### 3.2 Deep-learning model

#### 3.2.1 Trajectory gate recurrent unit

The gate recurrent unit (GRU) function [33] removes cell states and allows memory to directly and linearly accumulate on hidden states, in contrast to LSTM. The GRU controls the flow of information using update and reset gates. The update gate controls the update of the hidden state, and the reset gate determines whether to ignore the previous hidden state. GRU couples the input gate of LSTM with a forget gate, which can reduce the number of parameters and computational cost without significantly degrading network performance. Some researchers have suggested that a fully connected GRU encodes only temporal information and not spatial information [30]. To solve this problem, convolutional recurrent neural networks have been proposed. Replacing full connection with a convolution operation has been recommended to achieve transformation from input to state and state to state. The convolutional gate recurrent unit (ConvGRU) consists of the basic convolutional layer and the GRU layer. The kernel size and padding of the ordinary convolutional layer are fixed, and thus the points of the convolutional area are generally fixed. However, natural motion and changing wave fields exhibit position changes, and fixed points in a certain region may not be closely related; therefore, position-variant convolution replaces the ordinary convolutional layer, namely TrajGRU [18], the structure of which is illustrated in Fig. 5. The flow field $U_t$, $V_t$ is generated by the structure-generating sub-network $\gamma$ according to the input $x_t$ at the current time and the hidden state $h_{t-1}$ at the previous time. Then, the position designated by $U$ and $V$ is selected using the *warp()* function. Through these two operations, the current input and previous state are used to generate a local neighborhood set for each location at each time. The formulas for calculating each variable in the figure at time $t$ are as follows:

$$U_t, V_t = \gamma(x_t, h_{t-1}), \tag{1}$$

$$r_t = \sigma\left(\sum_{l=1}^{L} W_{rh}^l * warp(h_{t-1}, U_{t,l}, V_{t,l}) + W_{rx} * x_t + b_r\right), \tag{2}$$

$$z_t = \sigma\left(\sum_{l=1}^{L} W_{zh}^l * warp(h_{t-1}, U_{t,l}, V_{t,l}) + W_{zx} * x_t + b_z\right), \tag{3}$$

$$\tilde{h}_t = f\left(r_t \circ \left(\sum_{l=1}^{L} W_{\tilde{h}h}^l * warp(h_{t-1}, U_{t,l}, V_{t,l}) + W_{\tilde{h}x} * x_t + b_{\tilde{h}}\right)\right), \tag{4}$$

$$h_t = (1 - z_t) \circ h_{t-1} + z_t \circ \tilde{h}_t \tag{5}$$

where $U_t$, $V_t \in R^{L \times H \times W}$ is generated by the structure generation network $\gamma$, $L$ is the total number of local links at each position, and $f$ is the LeakyReLU activation function. The *warp()* function selects the position identified by $U$, $V$.

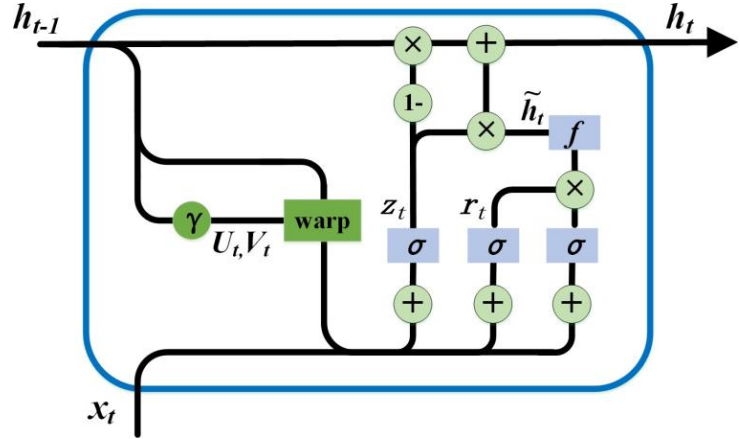

Figure 5: Structure of the TrajGRU cell. The rectangular box represents the neural-network layer, characters above the box represent the activation functions used by the neural network, and the circle represents the pointwise operation on the vector.

### 3.2.2 Structure of the TrajGRU model

Previous studies have shown that upper-ocean circulation in the North Pacific Ocean is dominated by the Black Current, a powerful western-boundary current that transports warm equatorial water toward the polar region [34]. Thus, changes in the spatial positions of wave fields are affected not only locally but also by a wide range of ocean circulation features. In addition, the WNP region is often impacted by strong typhoons and huge surface waves [35], around which the distribution of SWH shows dramatic changes and exhibits changes in track related to the movement of typhoons. Therefore, we use TrajGRU as the deep-learning model in the corrected method, employing a position-variant convolution process that can dynamically determine the evolutionary trajectory of SWH at each position.

The basic structure of the TrajGRU model is an encoder–decoder structure. The encoder component is implemented via stacking of down-sampling modules and TrajGRU units; the decoder component is implemented via stacking of up-sampling modules and TrajGRU units. These down-sampling modules and up-sampling modules, designated the downconv module and upconv modules, respectively, are deployed in the middle of the TrajGRU unit. These modules are composed mainly of a convolution layer and an activation layer. The network employs a three-layer encoder–decoder structure. In the encoder, the down-sampling module extracts feature information and transmits it to the TrajGRU unit of the next layer; the TrajGRU unit recurrently generates the hidden state at each moment as input timing information, which is passed to the next down-sampling module (purple arrow in Fig. 6). The hidden state at the last moment is passed to the TrajGRU unit of the corresponding layer in the decoder (red arrow in Fig. 6); meanwhile, the decoder connects the layers in the opposite direction and uses deconvolution to achieve final output prediction for the up-sampling operation.

The high-level states of the deep network are able to cover the global spatiotemporal scale, which guides the update of low-level states and thus further influences predictions. Our configuration used an input sequence of SWH predictions of length 5, an output sequence of length 1 containing correction results, and an input image size of $180 \times 300$. An architectural diagram of the model is shown in Fig. 6.

### 3.3 Pixel-switch loss function

In this section, we discuss how the pixel-switch loss function is mathematically derived for the SWH correction task and the reasons for adopting this method.

SWH forecast bias correction is a regression task, and therefore all models used in this study were regression models. The loss function of a regression model measures the gap between the output of the model and ERA5. In general, the mean square error (MSE) and mean absolute error (MAE) were used as loss terms for the regression task (Eq. (6)). The input patch of size $180 \times 300$ contains some land, and updating of the network parameters would be affected by these data; therefore, we generated a corresponding static mask for the land area of each image. The mask was used to systematically exclude pixels distributed over land when calculating the loss function during training. In addition, the weight value of the masked pixels was set to zero (Eq. (7)).

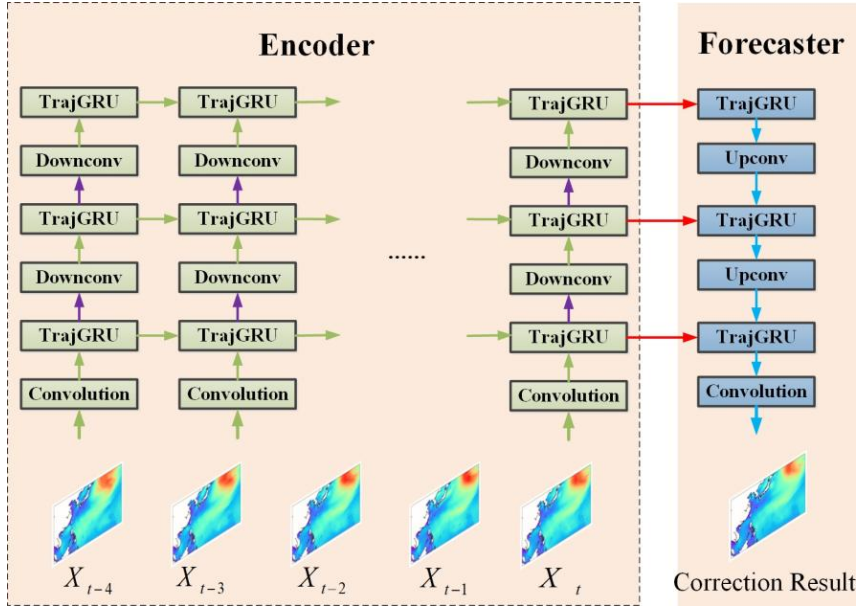

Figure 6: Architecture of TrajGRU. The encoding process is conducted by the down-sampling convolution module (downconv) and TrajGRU unit stacked on the left, and the decoding process is conducted by the up-sampling convolution module (upconv) and TrajGRU unit stacked on the right. Green arrows represent skip connections, which preserve fine-grained information from shallower layers for better prediction.

$$loss = \frac{1}{L} \sum_{b=1}^{B} \sum_{i=1}^{H} \sum_{j=1}^{W} \left[ \left| x_{bij} - \tilde{x}_{bij} \right| + (x_{bij} - \tilde{x}_{bij})^2 \right] \tag{6}$$

where $L = B \times H \times W$ . $B$ is the number of samples in each batch, and $B = 4$ in this study. $H$ and $W$ are the height and width of the model output, with values of 180 and 300 in this study, respectively. $\tilde{x}$ represents the ground-truth data and $x$ the model output.

$$loss = \frac{1}{O} \sum_{b=1}^{B} \sum_{i=1}^{H} \sum_{j=1}^{W} Mask_{bij} \left[ \left| x_{bij} - \tilde{x}_{bij} \right| + (x_{bij} - \tilde{x}_{bij})^2 \right] \tag{7}$$

where $O$ is the number of pixels representing the ocean area in each batch; the remaining terms are the same as in Eq. (6).

After the pre-training process, the validation set was used for evaluation to obtain the optimal model. Although the performance of the deep-learning model was better than that of XGBoost at correction, room for improvement in the accuracy of the corrected product remains. Learning architectures typically employ RMSE and MAE (Eq. (6)) as loss functions for regression tasks. The inherent averaging effect of these traditional loss functions enables deep-learning architectures to handle uncertainty by smoothing predictions [19]. This mechanism typically leads to the conditional bias problem in the model [36], which manifests as underestimation of the peak SWH value and overestimation of low SWH values, as shown in Fig. 7. In other words, the bias between the forecasts and actual observations or reanalysis data may be large. When training using the traditional loss function, tiny numbers of pixels with large biases are affected by massive pixels with small biases [19, 36]. In this manner, although the traditional loss function will gradually decrease during the training process, it will optimize the massive pixels with small biases rather than reduce the tiny number of pixels with large biases that we focus on.

Large-bias pixels correspond to the following two cases. First, they may be ECMWF high-wave misses: SWH in the sea area under abnormal weather conditions at time $t$ is often very large, and the waves reach the level of high waves (SWH > $4m$). If the ECMWF forecast for time $t$ does not show high waves in the area, a marine disaster may occur. Therefore, correcting these pixels is of economic value and can prevent casualties. These pixels are difficult to correct using traditional loss functions due to their tiny number. Second, these pixels may be ECMWF high-wave false alarms:

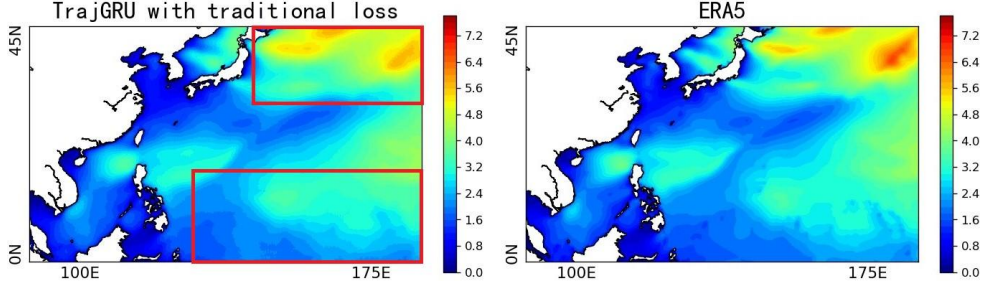

Figure 7: The problem of conditional bias for deep-learning models. The left panel shows the corrected product of the deep-learning model trained using the traditional loss function, and the right panel shows the ground truth ERA5 data at the corresponding moment. Large values of SWH are underestimated, while small values are overestimated (red boxes surround these areas).

the SWH of a sea area under normal weather at time $t$ is often small, and high waves should not be present. In this case, if the ECMWF forecast shows a high wave, and the ground-truth data indicate that the wave is small, such pixels may be ignored due to their tiny number. If such pixels are located in a major waterway, a false alarm may cause large economic losses.

To mitigate conditional bias, we propose the pixel-switch loss function. This function dynamically generates a pixel-switch matrix, forcing the network always to focus on those pixels with large biases. To obtain this loss function, we construct the characteristic function of the set [37] for pixels with large errors between the corrected model output and ground-truth data.

Let $A$ be the set of pixels with large error (defined as $e_{bij}$ in Eq. (10)) in a sample and $\delta_A(x)$ be the characteristic function of set $A$, which is defined as follows:

$$\delta_A(x) = \begin{cases} 1 & x \in A \\ 0 & x \notin A \end{cases} \tag{8}$$

From the above Eq. (8), $\delta_A(x)$ can map the set $A$ to $\{0, 1\}$; obtaining the 0,1 data can be considered a pixel switch.

The pixel-switch loss function is implemented as follows: first, we compute the average error for pixels between the model output and the ground-truth data for each sample, and the error is evaluated using $MAE + MSE$:

$$E_b = \frac{1}{S} \sum_{i=1}^{H} \sum_{j=1}^{W} Mask_{bij} \left[ \left| x_{bij} - \tilde{x}_{bij} \right| + (x_{bij} - \tilde{x}_{bij})^2 \right] \tag{9}$$

where $x_{bij}$ is the model output of the current sample $b$, $\tilde{x}_{bij}$ is the ground-truth data for the current sample $b$, and $E_b$ represents the average error of the pixels in sample $b$. $S$ is the number of pixels in the study area in sample $b$. $H, W$ are the same as in Eq. (6). Note that $E_b$ is sample-specific.

The corresponding threshold $\mu_b$ is set according to the $E_b$ of each sample: $\mu_b = \gamma E_b$. $\gamma$ is a constant, which is set empirically to $1/2$ in this study. The switch for pixels with an error above the threshold $\mu_b$ is turned on, and the corresponding element in the pixel-switch matrix is 1. The switch of a pixel with an error below $\mu_b$ is turned off, and the corresponding element in the pixel-switch matrix is 0.

The dynamic pixel switch matrix $\delta$ is calculated using the following pixel-switch loss function:

$$\delta_{bij} = \frac{sgn(e_{bij} - u_b) + sgn(|e_{bij} - u_b|)}{2} = \begin{cases} 1, & e_{bij} > u_b \\ 0, & e_{bij} \leq u_b \end{cases} \tag{10}$$

$$e_{bij} = \left| x_{bij} - \tilde{x}_{bij} \right| + (x_{bij} - \tilde{x}_{bij})^2$$

where $sgn()$ represents the sign function and is defined as follows:

$$sgn(x) = \begin{cases} -1, & x < 0 \\ 0, & x = 0 \\ 1, & x > 0 \end{cases} \qquad (11)$$

By comparing the error $e_{bij}$ at each pixel with the threshold $\mu_b$ in an element-wise manner, when $e_{bij} > u_b$, the element is considered to belong to set $A$, and the corresponding element in $\delta$ is set to 1. When $e_{bij} < u_b$, the pixel is considered not to belong to set $A$, and its element in $\delta$ is set to 0. Note that $E_b, \mu_b$ and $e_{bij}$ are all related to sample $b$, so $\delta_{bij}$ is the sample-specific pixel-switch term.

Finally, the pixel-switch loss function is obtained via pixel-wise multiplication of the pixel-switch matrix and the static mask loss function, as follows:

$$loss_{pixel-switch} = \frac{1}{O}\sum_{b=1}^{B}\sum_{i=1}^{H}\sum_{j=1}^{W} \delta_{bij} Mask_{bij}\left[\left|x_{bij} - \tilde{x}_{bij}\right| + (x_{bij} - \tilde{x}_{bij})^2\right] \qquad (12)$$

In this study, the pixel-switch loss function is proposed, which is essentially a weighted loss function. The loss weight of a pixel with an error greater than the threshold is set to 1, and the loss weight of a pixel with an error less than the threshold is set to 0. This loss function provides flexibility to adjust the training process dynamically and mitigates conditional bias by forcing the network to focus on specific high-error cases at the pixel level, as shown in Fig. 8.

As shown in the first row of Fig. 8, the positions of pixel switches are dynamically turned on and always applied to pixels with relatively large gaps or biases. The second row in Fig. 8 indicates that as the number of epochs increases, the gaps (red area within the black box) decrease significantly. This change implies that the pixel-switch loss function allows the TrajGRU model used in this study to be optimized rapidly by focusing on pixels with large gaps during training.

### 3.4 Pre-training and fine-tuning

Pre-training of deep-learning models has recently become a major research topic in computer vision and image processing [39, 40, 41, 42]. Pre-training of a deep-learning model refers to training the model for related tasks to help the model to learn parameters useful for its target tasks. After pre-training, the model is fine-tuned for the downstream target tasks. Pre-training models have also been applied recently to ocean parameter prediction; for example, historical climate model outputs (e.g., CMIP6 data [43]) have been used to pre-train the residual network [44], and reanalysis data (such as ERA5 data) have been applied to further optimize the forecast results for the following 6 h to 5 days. The results have shown that pre-training with climate model outputs prevents overfitting and improves performance in forecasting tasks [45]. Notably, those models require input of reanalysis data when they are used for correction and inference, and such data are difficult to obtain in real time, requiring time-delay methods or post-correction. Inspired by the studies noted above, this paper proposes a correction method based on a pre-training model. The differences from previous studies are as follows. We used the same input at the pre-training stage and the fine-tuning stage but adopted a new loss function (pixel-switch loss function) to fine-tune the specific downstream task. Furthermore, only ECMWF forecast data were used to construct the input features, mitigating the need for delayed reanalysis data and allowing real-time correction to be accomplished. The experimental results show that the proposed correction method has marked advantages. Implementation of this method is illustrated in Fig. 9. Above the dashed line is the pre-training process: the traditional loss function (Eq. (6)) is used to train a deep-learning model (e.g., TrajGRU), and preliminary evaluation allows selection of the corrected model that performs best on the validation set. Below the dotted line is the fine-tuning process: initialization is conducted using the parameters of the optimal correction model. The objective of this stage is to reduce the errors in specific pixel points, and therefore the pixel-switch loss function (Eq. (12)) is used for fine-tuning, after which the product is further optimized and corrected at the pixel level.

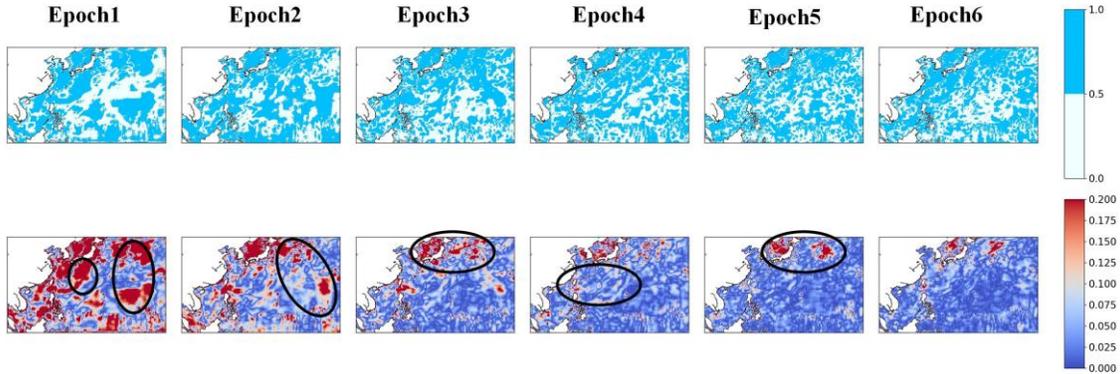

Figure 8: The effect of the pixel-switch term $\delta_{bij}$: turn on the switch of pixels with large errors, turn off the switch of pixels with small errors, and reduces the large errors. Epoch in this figure is defined as all the data in the training set used to conduct complete training of the model, which is designated as one epoch [38]. As the number of epochs increases, the first row provides visualization of the dynamic pixel-switch, with blue representing on, and white representing off. The second row provides visualization of the difference between the model output and the ground-truth data (ERA5), with red area indicating large differences.

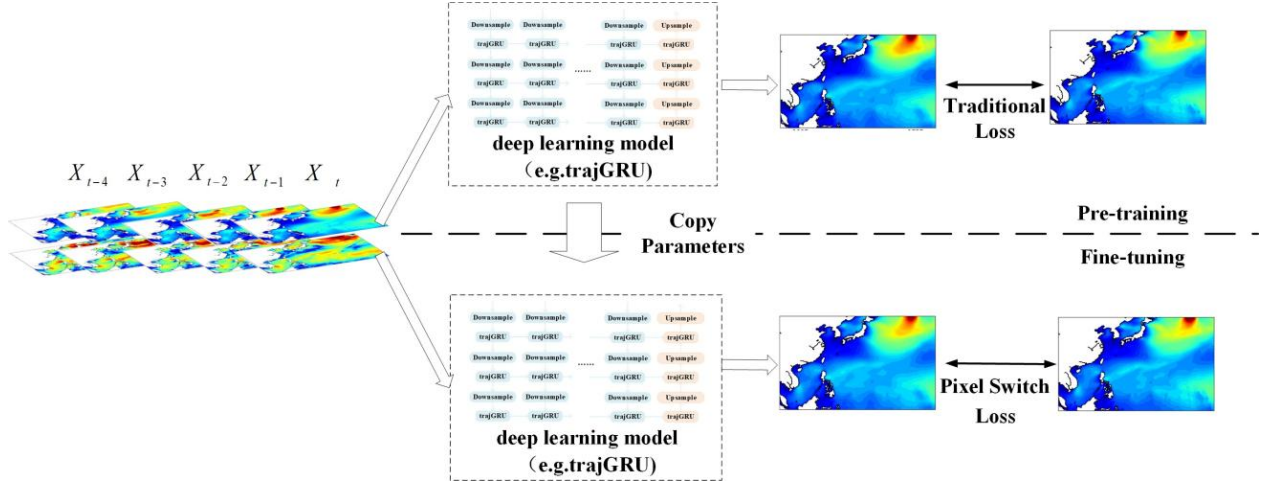

Figure 9: Pre-training and fine-tuning correction based on TrajGRU. The training procedure consists of a pre-training process (top) and a fine-tuning process (bottom).

## 3.5 Validation method

The MAE is the average error between the predicted and observed values. In general, a smaller MAE is better. MAE is calculated as follows:

$$MAE = \frac{1}{N \times L \times W} \sum_{n=1}^{N} \sum_{i=1}^{L} \sum_{j=1}^{W} |S_{nij} - O_{nij}| \tag{13}$$

RMSE is the square root of the ratio of the square of the deviation between the predicted and observed values to the number of observations. RMSE is used to measure the deviation between an observed value and the corresponding predicted value, and is one of the performance-evaluation indicators most commonly applied to regression problems. In general, a smaller RMSE is better. The RMSE Equation is as follows:

$$RMSE = \sqrt{\frac{1}{N \times L \times W} \sum_{n=1}^{N} \sum_{i=1}^{L} \sum_{j=1}^{W} (S_{nij} - O_{nij})^2} \tag{14}$$

where $N$ represents the number of samples in the test dataset; $L$ represents the length of the study area on the $x$-axis; and $W$ represents the width of the study area on the $y$-axis. $S$ represents the predicted or corrected value at $(i, j)$ at the forecast release time $t$, and $O$ represents the observed value at $(i, j)$.

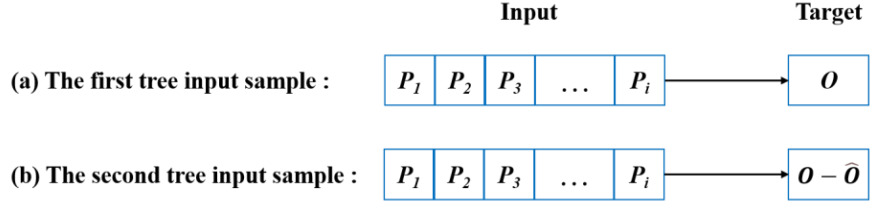

Figure 10: Input sample for XGBoost.

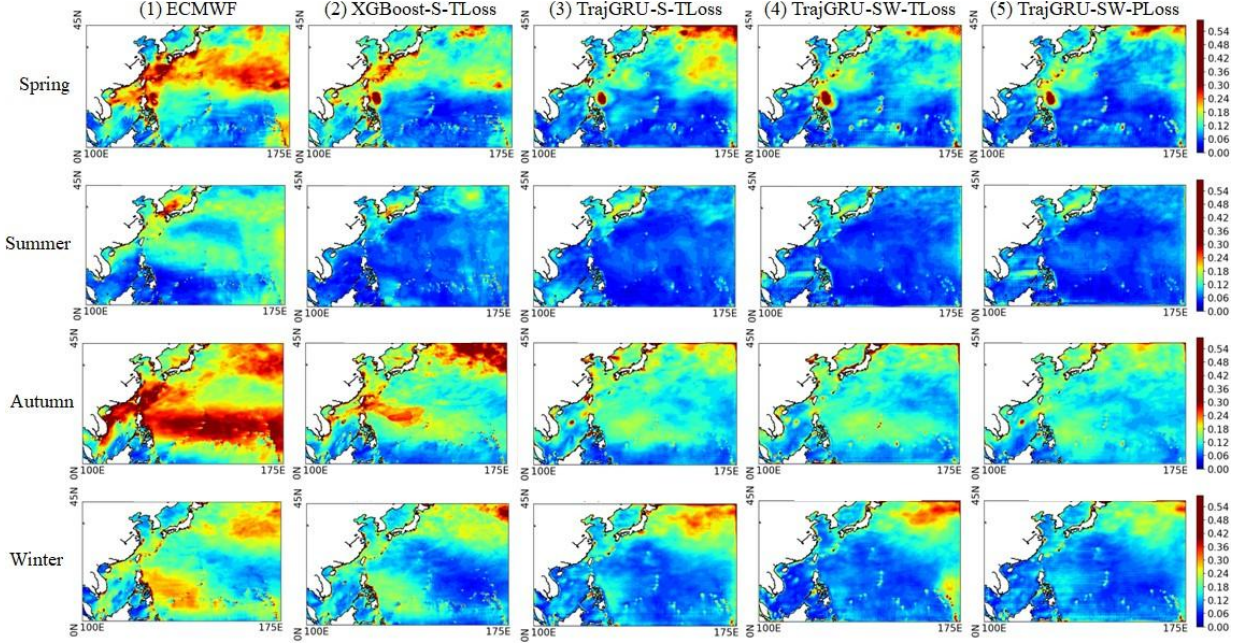

Figure 11: Root mean square error (RMSE) distribution of the 18-h lead time forecast after SWH correction in the various seasons of 2021. Column (1) is the forecast error in ECMWF, column (2) is the corrected product based on XGBoost, column (3) is the SWH-driven corrected product for TrajGRU, column (4) is the modified product with SWH and wind speed as drivers for TrajGRU, and column (5) is the modified product with SWH- and wind-speed-driven TrajGRU as well as fine-tuning of the correction function.

# 4 Experiments and Analyses

## 4.1 XGBoost method

As noted above, XGBoost is widely used in ocean-parameter correction and forecasting. Therefore, this study used XGBoost as a benchmark for comparison with the performance of the proposed deep-learning-based correction method [46, 47, 48]. The principle of XGBoost is briefly described in this section.

The ensemble method of XGBoost is a boosting ensemble that concatenates individual decision trees, with each decision tree depending on the result of the previous decision tree. In the XGBoost algorithm, to reduce the error in the training results from the previous decision tree, the target value of the next tree is not the true value but the error in the previous tree. The integration steps for this experiment are:

1. Suppose a sample consists of multiple numerical forecast member values $P_i$ and observations $O$. The first decision tree is trained using this sample to produce a predicted value $\hat{O}$ (Fig. 10a).

2. During training of the second decision tree, the input samples are multiple numerical prediction member values $P_i$ and observation errors $O - \hat{O}$ (Fig. 10b), and the input of the subsequent decision tree is the error in the previous decision tree; this step is iterative.

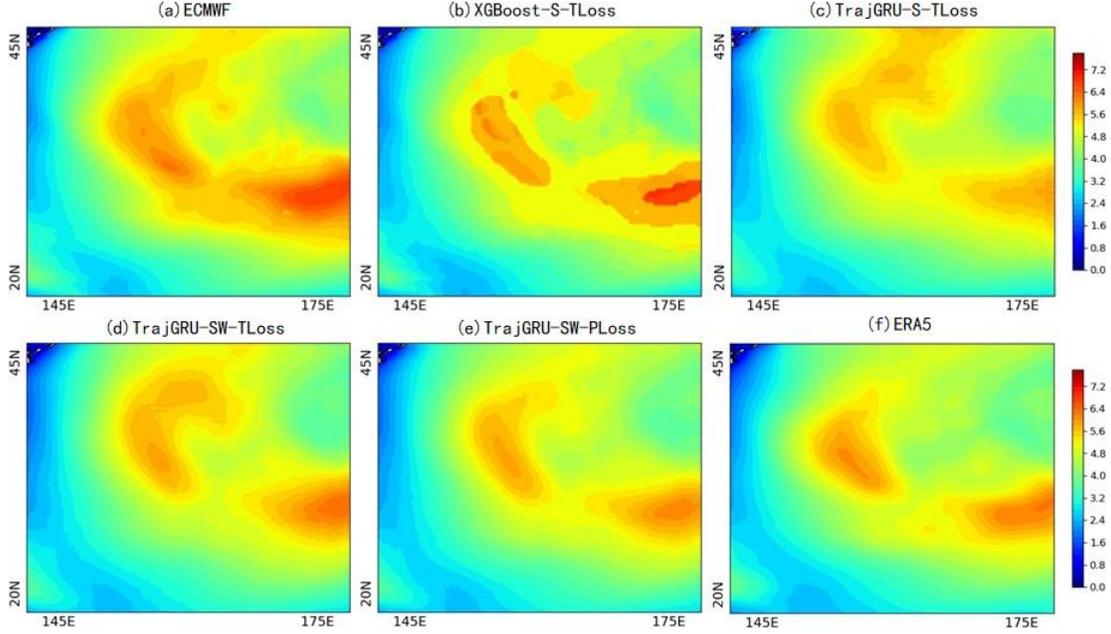

Figure 12: The 18-h forecast at 12:00 UTC on February 13, 2021 and 18-h lead time forecast SWH correction results: (a) ECMWF; (b) using the XGBoost model for correction; (c) using the TrajGRU model with only SWH input; (d) using the TrajGRU model with SWH and wind-speed inputs; (e) fine-tuning using the pixel-switch loss function based on (d); and (f) ERA5.

## 4.2 Correction results for SWH

The RMSE of the ECMWF-IFS forecast data is shown in column (1) of Fig. 11. Red indicates a large RMSE, and blue indicates a small RMSE. Column (1) shows that errors in spring, autumn, and winter were large, while the prediction error in summer was minimal. From the perspective of spatial distribution, errors were relatively large in the middle and high latitudes of the WNP.

TrajGRU-S-TLoss, TrajGRU-SW-TLoss, and TrajGRU-SW-Ploss (columns (3), (4), and (5)) performed better than XGBoost-S-TLoss (column (2)) in all seasons, especially in autumn. This implies that the deep-learning model has a much greater ability to correct SWH compared with traditional machine-learning methods. Moreover, the dark blue area of the TrajGRU-SW-TLoss model was significantly increased compared with the TrajGRU-S-TLoss model, indicating that adding the wind field as a driver based on wave-field-driven prediction can allow the model to extract more features, thereby improving correction performance. Finally, TrajGRU-SW-PLoss (rightmost column) had a significantly lower RMSE than that of the original ECMWF-IFS, with maximum values in the dark red region ranging from 0.3 to < 0.1 (high latitudes in spring). In addition, the red and yellow area of the corrected product from this model was the smallest among the models in all seasons, indicating that TrajGRU-SW-PLoss has the best SWH correction performance in allfour seasons.

The corrected RMSE is higher in the Pacific region east of the Philippines than in other regions in spring because Typhoon Surigae formed on the ocean surface east of the Philippines on April 14, 2021 and then moved northward. The intensity of the typhoon continued to increase, reaching its peak in the evening of April 17. The wind strengthened to above level 17, 68 $m/s$, and the pressure dropped to 905 $hPa$. Typhoon Surigae remained until April 23, 2021.

Fig. 12 shows a comparison of the corrected products for the 18-h lead-time forecast at 00:00 UTC on February 13, 2021. The results show that the outputs corrected using TrajGRU are more consistent with the observations (ERA5) compared with other model products. Notably, ERA5 data are reanalysis data, which are smoother than ECMWF forecast and XGBoost correction data [49]. As TrajGRU used ERA5 data as the ground-truth data for correction, the results tended to be continuous and smooth. The corrected product of the TrajGRU-SW-PLoss model was optimal; specifically, TrajGRU-SW-PLoss accurately corrected the area of large waves in the figure. Compared with the original ECMWF forecast and TrajGRU-SW-TLoss model, the contour of the large-wave distribution from TrajGRU-SW-PLoss showed the best consistency with ERA5. This indicates that compared with other methods the correction product after fine-tuning using the pixel-switch loss function better recovered the details lost due to the presence of large error points.

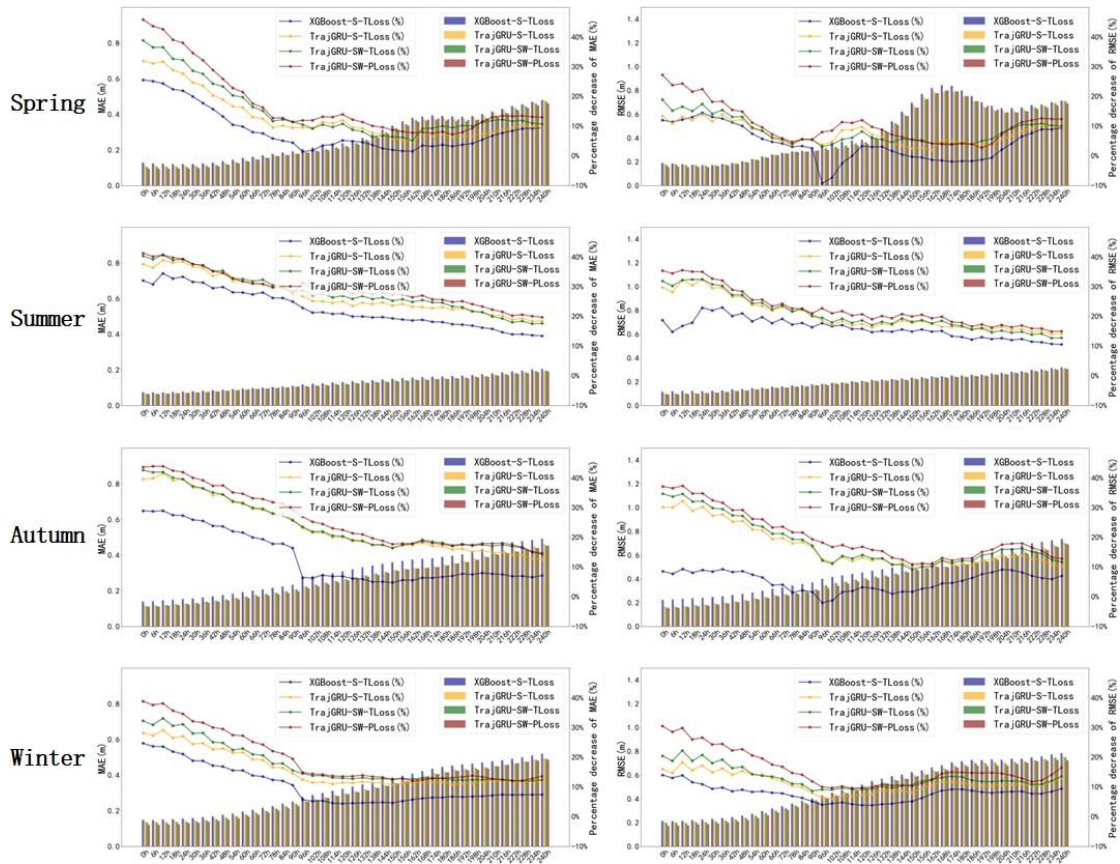

Figure 13: Variations in the RMSE and MAE for SWH predictions with lead times of 0–240 h after correction. The vertical axis on the left represents the MAE (RMSE) of the corrected product. The vertical axis on the right represents the percentage decrease in the MAE (RMSE) after correction. The horizontal axis represents the forecast lead time. The curve represents the percentage decrease in the RMSE or MAE after correction, with larger values being better. The histogram represents the corrected RMSE or MAE values, with smaller values being better.

For longer-term forecast products of SWH, Fig. 13 shows the difference in RMSE among the four corrected models for various forecast lead times (0–240 h). In the figure, the decreased curves of RMSE and MAE show apparent breaks at 96 h, due to time-segment correction based on different SWH data; the two correction models were trained for SWH with predicted lead times of 0–90 h and 96–240 h to improve the correction ability of the model while reducing the number of models as much as possible. The RMSE values were lower for products corrected using TrajGRU-S-TLoss, TrajGRU-SW-Tloss, and TrajGRU-SW-PLoss compared with XGBoost-S-TLoss. Compared with TrajGRU-S-TLoss, the RMSE of TrajGRU-SW-TLoss decreased by a further 1–5%, indicating that wind speed is an important parameter affecting SWH prediction. Compared with the single input variable of SWH, the combination of wind speed and SWH was a better input for the corrected model. TrajGRU-SW-PLoss achieved the smallest RMSE among all prediction lead times, indicating that the pixel-switch loss function designed in this paper effectively reduced the number of pixels with large errors. On the whole, with increasing forecast lead time, the percentage decreases in RMSE and MAE were reduced. Even for corrected products with a 240-h forecast lead time, the TrajGRU-SW-PLoss model had the strongest correction ability, which was most apparent in spring and winter. In spring, the MAE of TrajGRU-SW-PLoss-corrected products decreased by 13%, while the MAE of the other models decreased by only approximately 10%. In winter, the RMSE decreased by 17% after TrajGRU-SW-PLoss correction but decreased by only approximately 8–14% after correction using the other models.

To visualize the effect of SWH correction across different interval ranges and to further evaluate the performance of each correction model, scatter plots were constructed. The models for spring, summer, autumn, and winter were used to analyze the corrected mean with a 18-h forecast lead time. As shown in Fig. 14, the $x$-axis was the model output value, and the $y$-axis was the ground truth value (ERA5). The black line represents perfect reliability ($x = y$). The corrected product distribution of SWH is expected to be centered near the perfect reliability line, and a more compact distribution is better.

Table 2: Comparison of two types of correlation coefficients, CorrFE and CorrCE. Note that CorrFE is the correlation coefficient between the winter forecast data and ERA5, whereas CorrCE is the correlation coefficient between the winter corrected data and ERA5.

| CorrFE | CorrCE | | | |
|---|---|---|---|---|
| ECMWF | XGBoost-S-TLoss | TrajGRU-S-TLoss | TrajGRU-SW-Tloss | TrajGRU-SW-PLoss |
| 0.9862 | 0.9853 | 0.9865 | 0.9870 | 0.9898 |

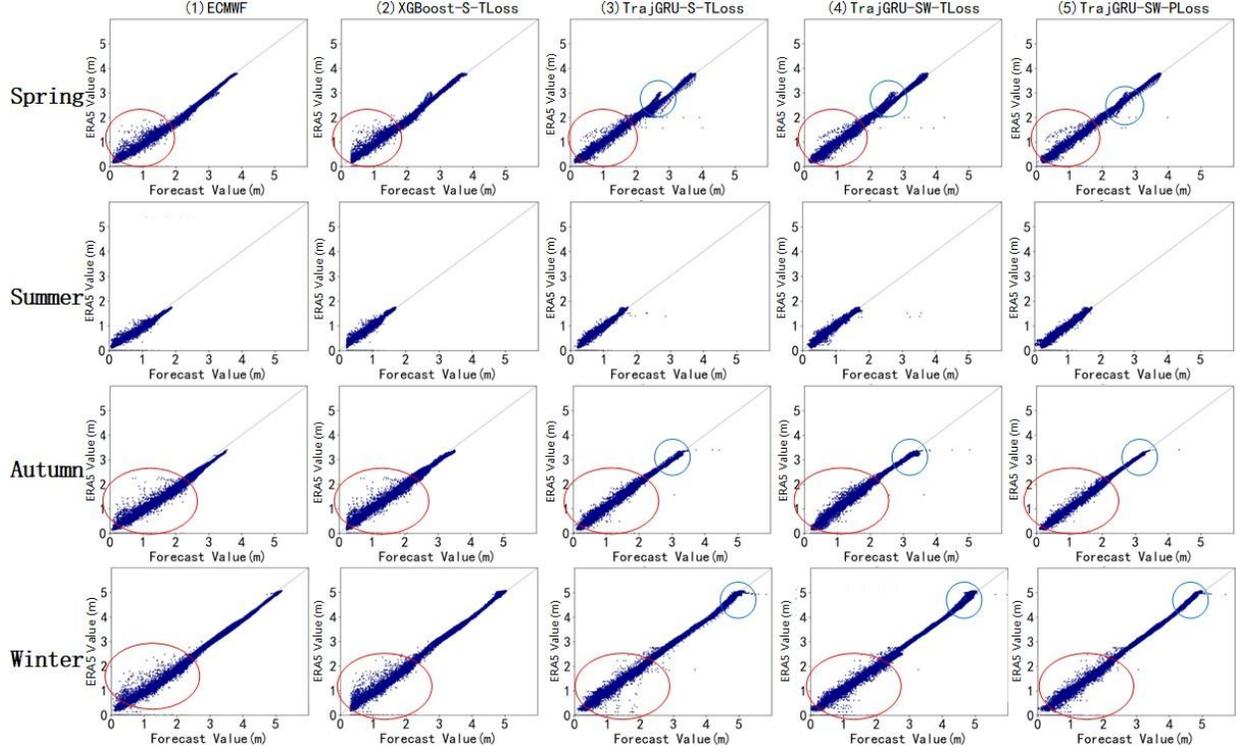

Figure 14: Scatter plot of the corrected 18-h lead-time forecasts in different seasons.

In Fig. 14, column (1) shows scatter plots of the forecast products from ECMWF, and columns (2)–(5) show scatter plots of products corrected using various correction models. This figure demonstrates that after correction of the ECMWF forecast product, the distribution of SWH was closer to the $x = y$ curve, especially in the circled portion of the figure. In the figure, red circles represent smaller waves and blue circles represent larger waves. These symbols indicate areas where significant differences occurred with correction. Over the whole year, SWH was concentrated in the range of 0–2.5 $m$ in summer, with the best correction results; the correction results for all four models were very close to the perfect reliability line. In spring, autumn, and winter, the XGBoost-S-TLoss model (column 2) was the least stable, with a weaker SWH correction effect in the interval of 0–2.5 $m$ than TrajGRU and a more diffuse distribution of SWH. Among the TrajGRU correction models (columns 3–5), the TrajGRU-SW-PLoss proposed in this study showed optimal performance, which was reflected in two aspects. First, for SWH in the interval 0–2 $m$, the corrected results of TrajGRU-SW-PLoss were closest to the $x = y$ curve, and the distribution was most concentrated among the models (red boxed area). Second, in the high SWH range, such as the 2.5–4 $m$ interval in spring and autumn and the 2.5–6 $m$ interval in winter, the SWH distribution was closest to the perfect reliability line. These results indicate that the predicted values with large errors are effectively corrected after fine-tuning using TrajGRU-SW-PLoss. The pixel-switch loss function used in TrajGRU-SW-PLoss for fine-tuning provided flexibility to adjust the training process and mitigate conditional bias by forcing the network to focus at the pixel level on specific cases with high errors.

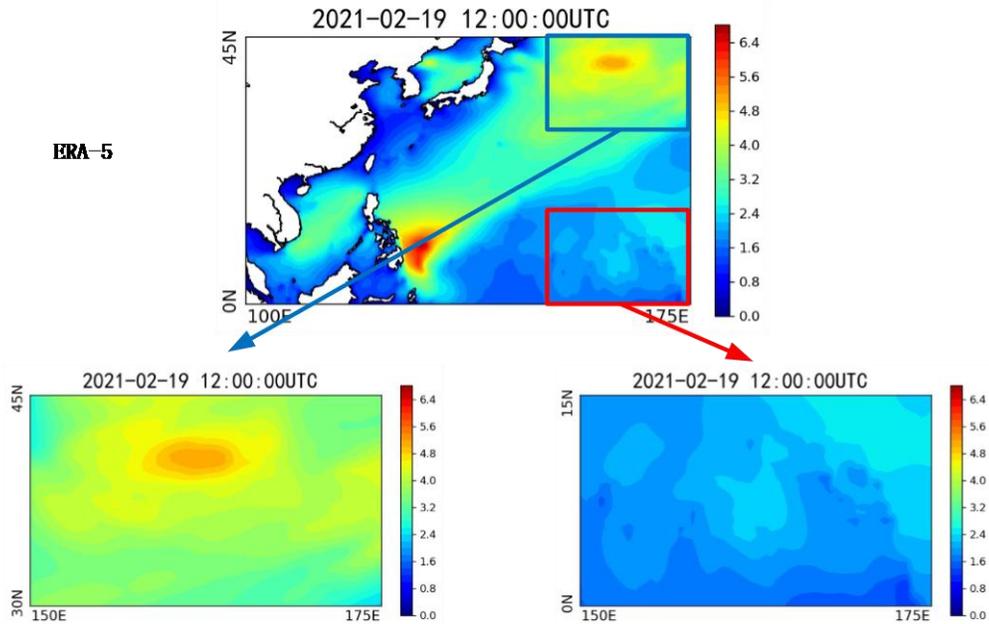

Figure 15: SWH visualization under extreme weather conditions (such as winter monsoons and tropical cyclones that cause huge waves) and normal weather conditions. The high-latitude sea area of 30–45 °N and 150–175 °E was used as analysis example $a$, representing extreme weather conditions (area circled in blue); the low-latitude area at 0–15 °N and 150–175 °E was used as example $b$, representing normal weather conditions (circled in red).

As shown in Table 2, forecasts corrected with the TrajGRU-SW-PLoss model have higher correlation coefficients (column 5 in CorrCE) with ERA5 data and are more similar to the ERA5 reanalysis data compared with forecasts from the ECMWF (CorrFEs) and forecasts corrected using other models (other CorrCEs).

Overall, SWH was concentrated in the range 0–6 $m$ in winter, reflecting large waves caused by extreme weather, which occur most frequently in winter. The SWH distribution in summer tended to be stable and was concentrated at 0–2 $m$, indicating mostly small to moderate wave sizes and rare extreme weather events.

### 4.3 Case study

Waves with a SWH greater than 2.5 $m$ are designated as rough seas. Accurate prediction of waves with a SWH greater than 2.5 $m$ can indicate the geographical distribution of sea areas with high incidences of extreme and disastrous weather processes, providing a reference for the selection and formulation of climate-adapted routes [50]. In addition, the extensive sea area in this region allows waves to grow easily to rough-sea size under the influence of a gale [50]. Therefore, this section discusses the correction results of TrajGRU-SW-PLoss for extreme SWH under both normal and extreme weather conditions to verify that the model has good correction ability for extreme SWH values.

We used the winter February 19, 2021 12:00:00 UTC sample as an example for analysis. Fig. 15 shows the ground truth (ERA5) data across the entire WNP at that time. SWH in the low-latitude sea area of 0–15 °N and 150–175 °E was between 0 and 2 $m$. Therefore, this area was used for analysis as an example of normal weather conditions (area enclosed in the red box). Meanwhile, SWH in the high-latitude sea area of 30–45 °N and 150–175 °E reached extreme values greater than 6 $m$. Therefore, this area was used for analysis as an example of extreme weather conditions (area enclosed in blue box). For comparative analysis, the ECMWF forecasts issued for February 19, 2021 at 12:00:00 UTC at a total of five issue times from February 15, 2021 at 00:00:00 UTC to February 19, 2021 at 12:00:00 UTC and the correction results for TrajGRU-SW-TLoss and TrajGRU-SW-PLoss at the same five issue times were used, as shown in Fig. 16. Compared with the corrected products in the TrajGRU-SW-PLoss model, listed in the three left columns for extreme weather conditions, the prediction results for ECMWF issued before February 18 were larger than the ground-truth data, and the predicted range of huge waves was larger. It took until February 18 for the forecast issued to approach the ground truth value. However, after TrajGRU-SW-PLoss correction, the area of large-wave occurrence overlapped almost perfectly with the area based on the ground-truth data, indicating that our correction method can improve the accuracy of the SWH forecast product to the reanalysis level 1 day in advance, which is of great significance for route formulation and selection. Notably, although very few pixels contained extremely large waves across the entire correction area, differences are apparent between the TrajGRU-SW-TLoss and TrajGRU-SW-PLoss correction results presented in the fourth and fifth rows of the figure, respectively. The three columns on the right represent comparisons between ECMWF forecast products and the TrajGRU-SW-TLoss and TrajGRU-SW-PLoss correction products under normal weather conditions. TrajGRU-SW-TLoss and TrajGRU-SW-PLoss both showed excellent correction ability for SWH values of 0–2 $m$. The corrected SWH distribution of the forecast product issued at 12:00 on February 17 was very similar to the ground-truth data, indicating that TrajGRU-SW-TLoss and TrajGRU-SW-PLoss introduced in this study can improve the accuracy of the SWH forecast product to the reanalysis level at least 2 days in advance.

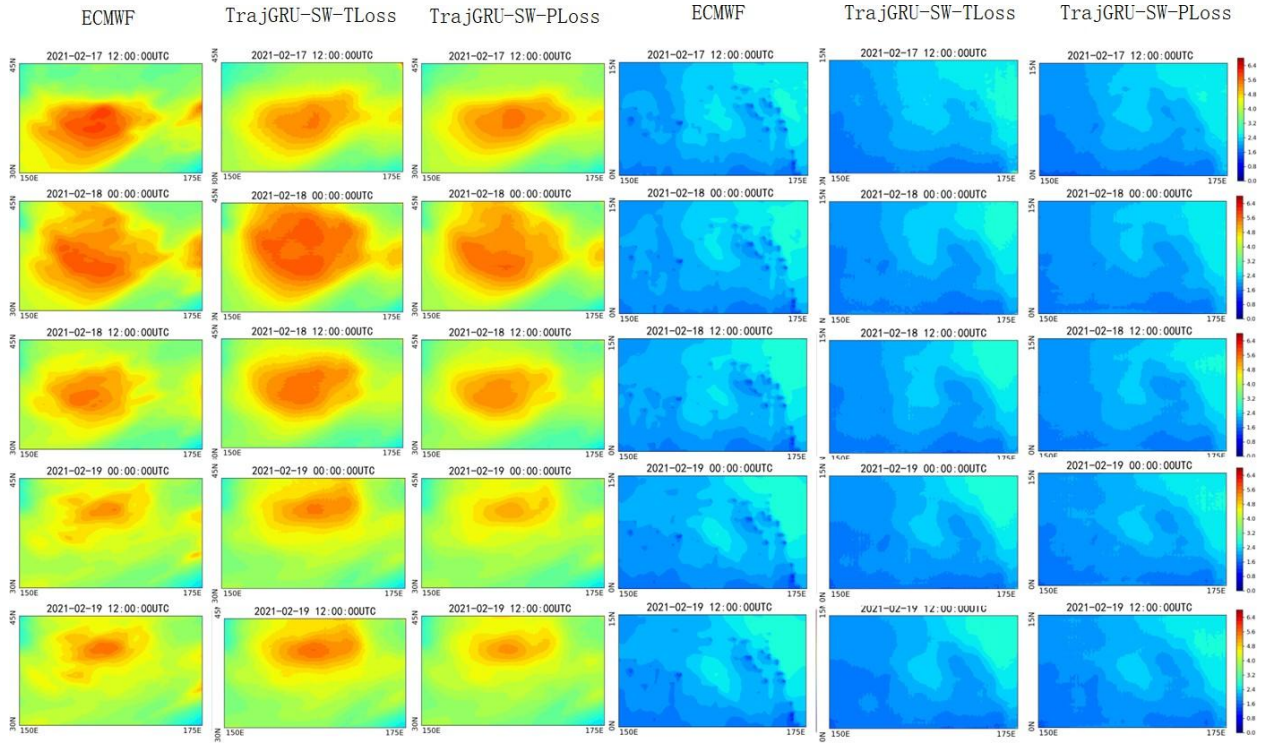

Figure 16: Visualization of ECMWF forecasts for February 19, 2021 at 12:00:00 UTC at issue times of 00:00:00 UTC on February 15, 2021 through 12:00:00 UTC on February 19, 2021, and correction results from the TrajGRU-SW-PLoss model for these five issue times. The sea areas under extreme and normal weather conditions are shown on the left and right, respectively. Although very few pixels contained extremely large waves across the entire correction area, differences are apparent between the TrajGRU-SW-TLoss and TrajGRU-SW-PLoss corrections presented in the fourthand fifth rows of the figure, respectively.

## 5 Conclusions

In this study, the forecast correction problem was transformed into a spatiotemporal sequence problem, and real-time rolling correction was developed for the 0–240$h$ ECMWF-IFS forecasts of SWH. The model was driven by both wave and wind fields, which effectively improved the correction results. The pixel-switch loss function, which fine-tuned the pre-trained correction model, further reduced RMSE and MAE by 2–8%. As a small number of pixels have large deviations, this result demonstrated the effectiveness of the pixel-switch loss function.

In the XGBoost method, each grid point remains independent during the correction process. However, ocean waves are continuous in both time and space (i.e., each grid point is affected by neighboring grid points). Therefore, spatiotemporal correlations must be considered, which is an advantage of TrajGRU, which learns from spatial information via convolution operations and computed optical flows between feature maps, as well as from temporal information via GRU units.

Notably, this study used the WNP as the study area; however, the bias correction method developed here is not limited to this area. In contrast to post-correction models requiring input of reanalysis data during correction, our model does not require input of reanalysis data, but instead directly maps the ECMWF-SWH forecast to ERA5 data, and it can be developed further as an operational real-time correction service. The model that maps wave forecast data to higher-precision reanalysis data in real time can be applied to any wave forecast data; it is not limited to ECMWF forecast data and ERA5 reanalysis data. These additional corrections will be explored in the near future at the National Marine Environmental Forecasting Center of China.

## Acknowledgment

This work was supported by National Natural Science Foundation (grant 42276202 and U1706218); and the National Key Research and Development Program of China (grant 2018YFC1407001). We want to thank *Qingdao AI Computing Center* and *Eco-Innovation Center* for providing inclusive computing power during the completion of this paper.